\documentclass[aps,pra,notitlepage,twocolumn,showpacs,superscriptaddress]{revtex4-1}
\usepackage{amsfonts}
\usepackage{}
\usepackage{amssymb}  
\usepackage{blindtext}
\usepackage{graphicx}  
\usepackage{dcolumn}   
\usepackage{amsmath,bm,amssymb,latexsym,galois,amsthm}   
\usepackage{dsfont} 
\usepackage{mathrsfs}
\usepackage[all,cmtip]{xy}
\usepackage{bm}

\usepackage{subfigure}
\usepackage{rotating}

\usepackage{ifpdf,xy}
\usepackage[braket]{qcircuit}
\usepackage{graphicx}

\def\bn{\begin{definition}}
\def\en{\end{definition}}
\def\ba{\begin{array}}
\def\ea{\end{array}}
\def\be{\begin{equation}}
\def\ee{\end{equation}}
\def\bd{\begin{description}}
\def\ed{\end{description}}
\def\bu{\begin{enumerate}}
\def\eu{\end{enumerate}}
\def\bi{\begin{itemize}}
\def\ei{\end{itemize}}
\usepackage[colorlinks,
            linkcolor=red,
            anchorcolor=red,
            citecolor=red,
            ]{hyperref}

\theoremstyle{plain}
\theoremstyle{definition}
\newtheorem{definition}{Definition}
\newtheorem{theorem}{Theorem}
\newtheorem{lemma}{Lemma}

\newtheorem*{exmp}{Example}
\newtheorem*{defn*}{Definition}
\def\eproof{{\mbox{}\hfill\qed}\medskip}

\theoremstyle{remark}

\newcommand{\Z}{{\mathbb{Z}}}
\newcommand{\R}{{\mathbb{R}}}
\newcommand{\C}{{\mathbb{C}}}

\renewcommand{\Re}{{\rm {Re}}}
\renewcommand{\Im}{{\rm {Im}}}
\def\ds{\displaystyle}

\begin{document}

\widetext


\title{\bf Building quantum neural networks based on swap test}
\author{Jian Zhao}
\affiliation{Key Laboratory of Quantum Information, Chinese Academy of Sciences, School of Physics, University of Science and Technology of China, Hefei, Anhui, 230026, P. R. China}
\affiliation{CAS Center For Excellence in Quantum Information and Quantum Physics, University of Science and Technology of China, Hefei, Anhui, 230026, P. R. China}
\author{Yuan-Hang Zhang}
\affiliation{School of the Gifted Young, University of Science and Technology of China, Hefei, Anhui, 230026, P.R. China.}
\author{Chang-Peng Shao}
\email{Email address: cpshao@amss.ac.cn}
\affiliation{Academy of Mathematics and Systems Science, Chinese Academy of Sciences, Beijing 100190, China}
\author{Yu-Chun Wu}
\email{Email address: wuyuchun@ustc.edu.cn}
\affiliation{Key Laboratory of Quantum Information, Chinese Academy of Sciences, School of Physics, University of Science and Technology of China, Hefei, Anhui, 230026, P. R. China}
\affiliation{CAS Center For Excellence in Quantum Information and Quantum Physics, University of Science and Technology of China, Hefei, Anhui, 230026, P. R. China}
\author{Guang-Can Guo}
\affiliation{Key Laboratory of Quantum Information, Chinese Academy of Sciences, School of Physics, University of Science and Technology of China, Hefei, Anhui, 230026, P. R. China}
\affiliation{CAS Center For Excellence in Quantum Information and Quantum Physics, University of Science and Technology of China, Hefei, Anhui, 230026, P. R. China}

\author{Guo-Ping Guo}
\email{Email address:  gpguo@ustc.edu.cn}
\affiliation{Key Laboratory of Quantum Information, Chinese Academy of Sciences, School of Physics, University of Science and Technology of China, Hefei, Anhui, 230026, P. R. China}
\affiliation{CAS Center For Excellence in Quantum Information and Quantum Physics, University of Science and Technology of China, Hefei, Anhui, 230026, P. R. China}
\affiliation{Origin Quantum Computing Hefei, Anhui 230026, P. R. China}


\begin{abstract}
Artificial neural network, consisting of many neurons in different layers, is an important method to simulate humain brain.
Usually, one neuron has two operations: one is linear, the other is nonlinear.
The linear operation is inner product and the nonlinear operation is represented by an activation function.
In this work, we introduce a kind of quantum neuron whose inputs and outputs are quantum states.
The inner product and activation operator of the quantum neurons can be realized by quantum circuits.
Based on the quantum neuron, we propose a model of quantum neural network
in which the weights between neurons are all quantum states.
We also construct a quantum circuit to realize this quantum neural network model.
A learning algorithm is proposed meanwhile.
We show the validity of learning algorithm theoretically and demonstrate the potential of the quantum neural network numerically.
\end{abstract}

\pacs{03.65.Ud, 03.67.Mn}
\maketitle

\section{Introduction}

Artificial neural networks can be traced back to McCulloch-Pitts (M-P) neurons
proposed in 1943 \cite{mcculloch1943logical}. Based on M-P neurons,
Rosenblatt in 1957  proposed the perceptron model with a learning algorithm \cite{rosenblatt1957perceptron}.
So far, artificial neural networks have had certain theoretical bases
\cite{minsky2017perceptrons,hospfield1982neural} and extensive practical
applications ranging from modeling, classification,
pattern recognition to multivariate data analysis \cite{basheer2000artificial,hippert2001neural}.

Quantum neural networks, proposed by Kak \cite{kak1995quantum} first in 1995, is a class
of neural networks that combine quantum information theory and artificial neural networks.
Different models related to quantum neural networks have been developed
\cite{schuld2014quest,altaisky2001quantum,andrecut2002quantum,da2016quantum,wan2017quantum,cao2017quantum,
farhi2018classification,daskin2018simple,toth1996quantum}.
Among these models, Ref. \cite{altaisky2001quantum} is a perceptron model with quantum input,
quantum output, and weights represented by operators, in which the concrete construction is not explained;
Ref. \cite{daskin2018simple} uses quantum computing to achieve the potential acceleration of classical neural networks;
Ref. \cite{toth1996quantum} is based on the actual physical device to construct an analog classical neural network.
However, there is still no uniform standard for the rigorous definition of quantum neural networks.

Recently, the paper \cite{shao2018quantum} introduced a strategy for using quantum phase
estimation to get the information for the inner product of two quantum states.
Inspired by this work, we introduce a definition of quantum neuron with quantum states
as input states, weights and a single-particle state as the output state.
Accordingly we propose a quantum neural network which can be represented by quantum circuits.
Besides, through theoretical analysis and the numerical experiment we demonstrate the validity of the learning algorithm.

Our starting point is to assume that there is a large amount of quantum states, each of
which is labeled by a quantum state. Given these data as the training set,
our goal is to predict the label of an unknown input state.
It is convenient for our proposed quantum neurons to process quantum data directly.
And it does not cost the classical computing resources to perform the trained quantum neural networks.
If using classical neural networks, one may need the method of quantum-state tomography to
reconstruct the quantum data \cite{lvovsky2009continuous}, which is a highly complex task itself.

This proposed neuron adapts to different kinds of data flexibly. When  quantum states
as the quantum data are labeled by real numbers rather than quantum states, we can
slightly modify the measured strategy to realize classical outputs. Things get more
complicated when both data and labels are classical. If using this proposed neuron we
need to consider the state preparation problem, which requires controlling the amplitude
of the desired quantum state to realize effectiveness \cite{soklakov2005state,soklakov2006efficient}.
A method making state preparation simple is to limit the structure of the data \cite{tacchino2018artificial},
in which they limit data to the vectors with binary value components.

The paper is organized as follows. At the end of this section we briefly state the notations used in this paper.
In section \ref{sec_swap test}, we describe the swap test and its quantum circuit.
In section \ref{sec_qn}, we construct a quantum neuron according to our proposed definition,
and then we analyze the property of this proposed quantum neuron.
The proof process is put in Appendix \ref{the proof of th1}.
In section \ref{sec_qnn}, based on the construction of quantum neuron we construct a kind of feedforward neural network
and a quantum circuit model representing the specific quantum neural network.
We give quantitative estimations of success probability and fidelity theoretically.
Some details are presented in Appendix \ref{the proof of thm2}.
We put the training process of the quantum neural network in section \ref{sec_training process}.
And in section \ref{sec_numer} we present an experiment for numerical simulation.
At last in section \ref{sec_conclusion}, we draw the conclusions of this paper.

\textbf{Notation.}
We use capital Roman letters $A$, $B$,$\ldots$, for matrices, lower case Roman letters $x$, $y$,$\ldots$, for column vectors,
and Greek letters $\alpha$, $\beta$,$\ldots$, for scalars. For a scalar $\alpha$,
we denote by {\bf Re}$\alpha$ and {\bf Im}$\alpha$ the real and imaginary part of $\alpha$, respectively.
Given a column vector $x$, $x^T$ denotes its transpose and
$x^{\dag}\triangleq(\bar x)^T$ is its conjugate transpose, and similar for a given matrix $A$.
Specifically, for the unitary transformation $U$, $U^{\dag}=U^{-1}$.
A quantum state $|x\rangle$ $\in$ $\C^{2^n}$ is regarded as the normalized vector.
We write
$R_Y(\beta)
=\left[
   \begin{array}{rr}
     \cos\frac{\beta}{2} &-\sin\frac{\beta}{2}\\
     \sin\frac{\beta}{2} & \cos\frac{\beta}{2} \\
   \end{array}
 \right]
$
and
$R_Z(\gamma)
=\left[
   \begin{array}{cc}
     e^{-\bm i\frac{\gamma}{2}} & 0\\
     0 & e^{\bm i\frac{\gamma}{2}} \\
   \end{array}
 \right].
$

\section{Swap test and its quantum circuit}\label{sec_swap test}
The swap test method has been applied widely to quantum machine learning \cite{lloyd2013quantum,lloyd2014quantum,zhang2016quantum}.
In this section, we describe the swap test and its quantum circuit.

Let $|x\rangle,|w\rangle\in \C^{2^n}$ be two quantum states
that are prepared by unitary operators $U_x,U_w$ respectively. That is
$|x\rangle=U_x|0\rangle^{\otimes n},|w\rangle=U_w|0\rangle^{\otimes n}$.
Swap test is a technique that can be used to estimate $\langle x|w\rangle$.
The basic procedure can be stated as follows:

{\bf Step 1.} Prepare the state
\be
|\phi_r\rangle=\frac{1}{\sqrt{2}}(|+\rangle|x\rangle+|-\rangle|w\rangle).
\ee
The quantum circuit to prepare $|\phi_r\rangle$ is simple; see Figure \ref{swap test:step 1} below.
We denote the unitary to prepare $|\phi_r\rangle$ as $U_{\phi_r}$.

\begin{figure}[h]
\hspace{2em}\Qcircuit @C=1em @R=.7em {
 \lstick{|0\rangle}& \qw &\gate{H} &\ctrlo{1} &\ctrl{1} &\gate{H} &\qw\\
 \lstick{|0\rangle^{\otimes n}}& \qw{/} & \qw & \gate{U_x} & \gate{U_w} &\qw&\qw
}
\hspace{2em}\Qcircuit @C=1em @R=.7em {
\lstick{}\\
\lstick{}
\inputgroup{1}{2}{1em}{\ket{\phi_r}}
}
\caption{Quantum circuit to prepare $|\phi_r\rangle$}
\label{swap test:step 1}
\end{figure}
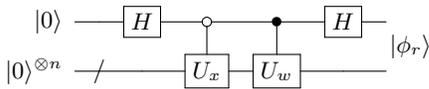

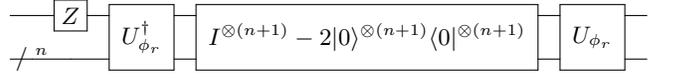
\begin{figure}[h]
\hspace{0.5em}\Qcircuit @C=0.9em @R=.7em {
 & \qw &\gate{Z} & \multigate{1}{U_{\phi_r}^{\dag}} & \multigate{1}{I^{\otimes (n+1)}-2|0\rangle^{\otimes (n+1)}\langle0|^{\otimes (n+1)}} & \multigate{1}{U_{\phi_r}} &\qw \\
 & \qw{/\hspace{0.3em}^n} & \qw & \ghost{U_{\phi_r}^{\dag}} & \ghost{I^{\otimes (n+1)}-2|0\rangle^{\otimes (n+1)}\langle0|^{\otimes (n+1)}} & \ghost{U_{\phi_r}} &\qw
}
\caption{Quantum circuit to implement $G_r$}
\label{swap test:step 2}
\end{figure}

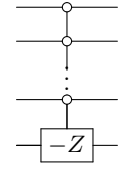
\begin{figure}[h]
\hspace{2.5em}\Qcircuit @C=1em @R=1em {
&\ctrlo{1}&\qw\\
&\ctrlo{1}&\qw\\
&\vdots&\\
&\ctrlo{1}\qwx[1]&\qw\\
&\gate{-Z}\qw&\qw
}
\caption{Quantum circuit to run $I^{\otimes (n+1)}-2|0\rangle^{\otimes (n+1)}\langle0|^{\otimes (n+1)}$}
\label{swap test:step 21}
\end{figure}

{\bf Step 2.} Construct the unitary transformation
\be\ba{lll} \vspace{.2cm}
G_r &=& (I^{\otimes (n+1)}-2|\phi_r\rangle\langle\phi_r|)(Z\otimes I^{\otimes n}) \\
&=& U_{\phi_r}(I^{\otimes (n+1)}-2|0\rangle^{\otimes (n+1)}\langle0|^{\otimes (n+1)}) U_{\phi_r}^\dag (Z\otimes I^{\otimes n}),
\ea\ee
where $Z=|0\rangle\langle 0| - |1\rangle\langle 1|$ is the Pauli-Z matrix.
The circuit to implement $G_r$ is represented in Figure \ref{swap test:step 2}.
As for the unitary operator $I^{\otimes (n+1)}-2|0\rangle^{\otimes (n+1)}\langle0|^{\otimes (n+1)}$,
we can run it in the circuit shown in Figure \ref{swap test:step 21}.

The state $|\phi_r\rangle$ can be rewritten as
\be
|\phi_r\rangle=\frac{1}{2}\left(|0\rangle(|x\rangle+|w\rangle)+|1\rangle(|x\rangle-|w\rangle)\right).
\ee
The amplitude of $|0\rangle$ is $\sqrt{1+{\bm \Re}\langle x|w\rangle}/\sqrt{2}$, and
the amplitude of $|1\rangle$ is $\sqrt{1-{\bm \Re}\langle x|w\rangle}/\sqrt{2}$.
Denote $|u\rangle$ and $|v\rangle$ as the normalized states of $|x\rangle+|w\rangle$ and $|x\rangle-|w\rangle$ respectively.
Then there is a real number $\theta_r\in[0,{\pi}/{2}]$ such that
\begin{align}\label{eq_phi_theta_0u1v}
|\phi_r\rangle=\sin\theta_r|0\rangle|u\rangle+\cos\theta_r|1\rangle|v\rangle.
\end{align}
Moreover, $\theta_r$ satisfies $\cos\theta_r=\sqrt{1-{\bm \Re}\langle x|w\rangle}/\sqrt{2}$, i.e.,
\be \label{swap test:inner product real part}
{\bm \Re}\langle x|w\rangle=-\cos{2\theta_r}.
\ee

Apply the Schmidt decomposition method to the quantum state $|\phi_r\rangle$, and we can decompose it into
\begin{align}\label{eq_phi_schmidt}
|\phi_r\rangle=\frac{-\bm i}{\sqrt{2}}(e^{\bm i \theta_r}|w_+\rangle-e^{-\bm i\theta_r}|w_-\rangle),
\end{align}
where
$
|w_{\pm}\rangle=\frac{1}{\sqrt{2}}(|0\rangle|u\rangle\pm\bm i|1\rangle|v\rangle).
$
Besides, it is easy to check that
\begin{align}\label{eq_Gw}
G_r|w_{\pm}\rangle=e^{\pm\bm i  2\theta_r}|w_{\pm}\rangle.
\end{align}
This means $|w_{\pm}\rangle$ are the eigenstates of $G_r$.
The information of $\theta$ is contained in the arguments of the eigenvalues.

\begin{figure}[h]
\hspace{-.7em}\Qcircuit @C=1em @R=.7em {\mbox{$ |j\rangle$}
}

\hspace{2em}\Qcircuit @C=1em @R=.7em {
 \lstick{|0\rangle^{\otimes t}}& \qw{/} &\gate{H^{\otimes t}}  &\ctrl{1} &\gate{{\rm{FT}}^{\dag}} &\qw\\
 \lstick{|\phi_r\rangle}& \qw{/} & \qw & \gate{G_r^j} &\qw&\qw
}
\hspace{2em}\Qcircuit @C=1em @R=.7em {
\lstick{}\\
\lstick{}
 \inputgroup{1}{2}{1em}{\ket{\psi_r}}
}
\caption{Quantum phase estimation to estimate $\theta$}
\label{swap test:step 3}
\end{figure}
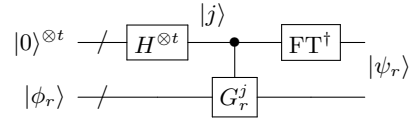

{\bf Step 3.} Use quantum phase estimation algorithm to estimate $\theta$.
The quantum circuit is shown in Figure \ref{swap test:step 3}.

In Figure \ref{swap test:step 3}, $t$ is an integer relates to the precision, and ${\rm FT}$ is the quantum Fourier transform.
The control gate $G^j_r$ should be regarded as a composition of a series of controlled gates $G^{2^{i}}_r$ by viewing the
$i$-th qubit in the first register as control qubit, where $i=0,\dots,t-1$.

By equations \eqref{eq_phi_schmidt} and \eqref{eq_Gw}, the output of quantum phase estimation is an approximate of
\begin{align}\label{eq_psi}
|\psi_r\rangle=\frac{-\bm i}{\sqrt{2}}(e^{\bm i \theta_r}|y_r\rangle|w_+\rangle-e^{-\bm i\theta_r}|2^t-y_r\rangle|w_-\rangle),
\end{align}
where $y_r\in[0,2^{t-1}]$ and $y_r\pi /2^{t-1}$ is an approximate of $2\theta_r$.
By equation (\ref{swap test:inner product real part}),
we have
\be \label{swap test:real part}
{\bm \Re}\langle x|w\rangle\approx -\cos (\pi y_r/2^{t-1}).
\ee

Note that
$
{\bm\Im}\langle x|w\rangle = - {\bm \Re}\langle x|\bm i|w\rangle,
$
thus the proposal to estimate the real part of inner product is also
suitable to estimate ${\bm\Im}\langle x|w\rangle$.
We only need to consider the state $|\phi_i\rangle=\frac{1}{\sqrt{2}}(|+\rangle|x\rangle-\bm i|-\rangle|w\rangle)$.
Finally, we will obtain a $y_i\in[0,2^{t-1}]$, such that
\be \label{swap test:imaginary part}
{\bm \Im}\langle x|w\rangle \approx -\cos{({\pi y_i}/{2^{t-1}})}.
\ee
For convenience, the corresponding parameters, unitaries and quantum states used to estimate ${\bm\Im}\langle x|w\rangle$
will be accordingly denoted
by $\theta_i, y_i, U_{\phi_i},G_i,U_{\psi_i}$ and $|\phi_i\rangle,|\psi_i\rangle$.

\section{Construction of the quantum neuron}\label{sec_qn}

\subsection{Definition of the quantum neuron}

A classical neuron
can be treated as a function that maps a vector $x=(x_1,\dots,x_n)^T\in\R^n$ to a real value
$z=f(x^T w)$, where $w=(w_1,\dots,w_n)^T\in\R^n$ and $f$ is usually a nonlinear function.
$\{x_i\}_{i=1}^{n}$ and $\{w_i\}_{i=1}^{n}$ are called the input values and synaptic weights, respectively.
The function $f$ is called the activation function.
Similarly, we propose the definition of quantum neuron as follows

\begin{definition}\label{def_qnn}
Let $\ket{w}=\ket{w_1,\dots,w_n}\in \mathcal (\C^2)^{\otimes n}$
be a product state. Denote $\mathcal B(0,1)=\left\{a\in \C : |a|\leqslant1\right\}$.
Assume that $f$ is a map from $\mathcal B(0,1)$ to the subspace of $\C^2$ with unit norm,
then the map
\be\ba{clc} \vspace{.2cm}
F: \C^{2^n}&\longrightarrow&  \C^{2} \\
  ~~|x\rangle &\longmapsto& f(\langle x|w\rangle)
\ea\ee
is called an $n$-variable quantum neuron.
\end{definition}

\begin{figure}
  \centering
  \includegraphics[width=8cm]{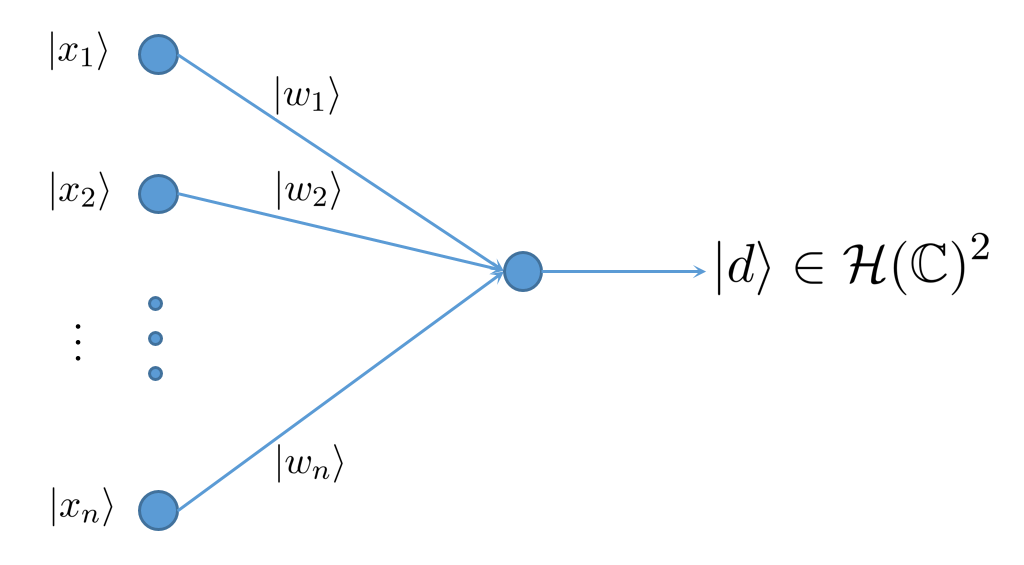}\\
  \caption{Structure of quantum neuron, where $|d\rangle = f(\langle x|w\rangle)$ is the output state.}
  \label{fig_qn}
\end{figure}

In the $n$-variable quantum neuron,
we call $|x\rangle$ the input state,
$\{|w_i\rangle\}_i$ the (synaptic) weight states and
$f(\langle x|w\rangle)$ the output state.
The map $f$ plays the role of activation function in defining the quantum neuron
Figure \ref{fig_qn} shows the basic structure of quantum neuron.

Assume that $a\in\C$, a commonly used activation function in this paper is
\be\ba{lll} \vspace{.2cm} \label{activtion function}
f(a) &=& \ds R_Z(-\frac{\pi}{2})R_Z(\arccos-{\bm \Im}a)R_Y(\arccos-{\bm \Re}a)|0\rangle \\
&=& \left[
      \begin{array}{l} \vspace{.2cm}
        \cos(\frac{\arccos-{\bm \Re}a}{2}) e^{\bm i (\frac{\pi}{4}-\frac{\arccos-{\bm \Im}a}{2})} \\
        \sin(\frac{\arccos-{\bm \Re}a}{2}) e^{-\bm i (\frac{\pi}{4}-\frac{\arccos-{\bm \Im}a}{2})}\\
      \end{array}
    \right].
\ea\ee
The operator $R_Z(-{\pi}/{2})$ is added to make sure that if $a\in\mathbb{R}$, then
\[
f(a) = \left[
      \begin{array}{l} \vspace{.2cm}
        \cos(\frac{\arccos-a}{2}) \\
        \sin(\frac{\arccos-a}{2}) \\
      \end{array}
    \right]\in \R^2.
\]

\subsection{Realization of the output state in the quantum circuit}
\label{subsec_ideal}

Now assume that the activation function $f$ is defined by equation (\ref{activtion function}).
Let $|x\rangle \in \C^{2^n}$ be an input state  and  $|w\rangle \in (\C^2)^{\otimes n}$ be a weight state.
In this subsection, we show how to realize $f(\langle x|w\rangle)$ in the quantum circuit.

We first show how to realize $f(\langle x|w\rangle)$ in the quantum circuit in the ideal case, then extend it into the general case.
By ideal, we mean both $\frac{\arccos-{\bm \Re}\langle x|w\rangle}{2\pi}$ and $\frac{\arccos-{\bm \Im}\langle x|w\rangle}{2\pi}$
can be represented in binary form with $t$ bits precisely.
As a result, swap test can approximate these two values with no error, i.e.,
equations (\ref{swap test:real part}) and (\ref{swap test:imaginary part}) are exact.

By equation (\ref{swap test:real part}), (\ref{swap test:imaginary part}) and (\ref{activtion function}),
\be \label{state-ideal}
f(\langle x|w\rangle)=R_Z(-{\pi}/{2}) R_Z({\pi y_i}/{2^{t-1}}) R_Y({\pi y_r}/{2^{t-1}}) |0\rangle,
\ee

To prepare the state (\ref{state-ideal}),
first we consider $|\psi_r\rangle|0\rangle$, where $|\psi_r\rangle$ is given in equation (\ref{eq_psi}).
We want to generate the state $R_Y({\pi y_r}/{2^{t-1}}) |0\rangle$ in the third register of
$|\psi_r\rangle|0\rangle$ by viewing $|y_r\rangle$ and $|2^t-y_r\rangle$ as control registers. That is
to obtain the following transformation
\[\ba{lll} \vspace{.2cm}
|\psi_r\rangle |0\rangle
&=& \ds \frac{-\bm i}{\sqrt{2}}(e^{\bm i \theta_r}|y_r\rangle|w_+\rangle-e^{-\bm i\theta_r}|2^t-y_r\rangle|w_-\rangle) |0\rangle \\
&\mapsto&  |\psi_r\rangle R_Y({\pi y_r}/{2^{t-1}})|0\rangle.
\ea\]

\begin{figure}[htbp]
\centering
\subfigure[ ]{
\begin{minipage}{10cm}
\hspace{-3em}\Qcircuit @C=0.4em @R=1em {
&&&& && \ctrl{6}&\ctrl{6} &\qw&\qw&\qw&\qw&\qw&\ctrl{6}&\qw&\\
&&&& &&\qw&\qw&\ctrl{5}&\qw&\qw&\qw&\qw&\qw&\qw&\\
&\qw{\hspace{0.4em}/\hspace{.2em}^t} &\qw&\qw& \scalebox{1.1}[1.6]{\Bigg\{}&&\qw &\qw &\qw&\qw&\qw&\qw&\qw&\qw&\qw&\\
&&&& && \cdots&&&&\ddots&&&&\\
&&&& && \qw&\qw&\qw&\qw&\qw&\qw&\ctrl{2}&\qw&\qw&\\
&\qw\hspace{1.4em}{/\hspace{0.2em}^{n+1}}&\qw &\qw&\qw \qw&\qw& \qw&\qw&\qw&\qw&\qw&\qw&\qw&\qw&\qw
\inputgroupv{3}{6}{.8em}{1.6em}{|\psi_r\rangle}&\\
\lstick{|0\rangle}&\qw&\qw&\qw& \qw&\qw&\gate{X}&\gate{R_Y(\pi)}&\gate{R_Y(\frac{\pi}{2})}&\qw&\cdots&&\gate{R_Y(\frac{\pi}{2^{t-1}})}&\gate{-X}&\qw&&&&|d_r\rangle
}
\end{minipage}\label{fig_qnideala}}
\subfigure[ ]{
\begin{minipage}{10cm}\scalebox{0.95}[1]{
\hspace{-3.5em}\Qcircuit @C=0.3em @R=1em {
&&&& && \ctrl{6}&\ctrl{6} &\qw&\qw&\qw&\qw&\qw&\ctrl{6}&\qw&\qw\\
&&&& &&\qw&\qw&\ctrl{5}&\qw&\qw&\qw&\qw&\qw&\qw&\qw\\
&\qw{\hspace{0.4em}/\hspace{.2em}^t} &\qw&\qw& \scalebox{1.1}[1.6]{\Bigg\{}&&\qw &\qw &\qw&\qw&\qw&\qw&\qw&\qw&\qw&\qw\\
&&&& && \cdots&&&&\ddots&&&&&\\
&&&& && \qw&\qw&\qw&\qw&\qw&\qw&\ctrl{2}&\qw&\qw&\qw\\
&\qw\hspace{1.4em}{/\hspace{0.2em}^{n+1}}&\qw &\qw&\qw \qw&\qw& \qw&\qw&\qw&\qw&\qw&\qw&\qw&\qw&\qw&\qw
\inputgroupv{3}{6}{.8em}{1.6em}{|\psi_i\rangle}&\\
\lstick{|d_r\rangle}&\qw&\qw&\qw& \qw&\qw&\gate{X}&\gate{R_Z({\pi})}&\gate{R_Z(\frac{\pi}{2})}&\qw&\cdots&&\gate{R_Z(\frac{\pi}{2^{t-1}})}&\gate{-X}&\qw&\qw&&&
}}
\end{minipage}\label{fig_qnidealb}}
\caption{The quantum neuron to generate $R_Z({\pi y_i}/{2^{t-1}}) |d_r\rangle$,
where $|d_r\rangle = R_Y({\pi y_r}/{2^{t-1}})|0\rangle$.}
\label{fig_qnideal0}
\end{figure}
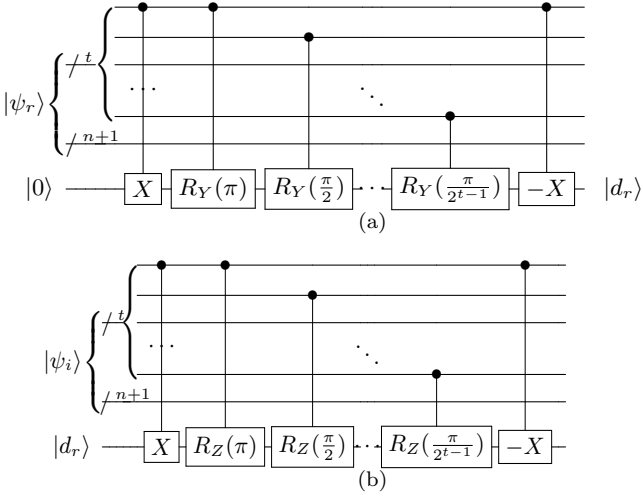

The control rotation generated by $|y_r\rangle$ gives $R_Y({\pi y_r}/{2^{t-1}})$ directly.
However, the control rotation generated by $|2^t-y_r\rangle$ gives
$R_Y(\pi (2^t-y_r)/2^{t-1}) = -X R_Y(\pi y_r/2^{t-1}) X$.
To modifies this, it suffices to add a control $X$ and control $-X$ gate.
More precisely, assume that $y_r'\in\{y_r,2^t-y_r\}$ and  $y_r'= \sum_{j=0}^{t-1} 2^{j} y'_{r,t-j-1}$ in binary form, then
the control qubit is $|y'_{r,0}\rangle$.
If $y'_{r,0}=0$, then we know $y_r'=y_r$ and we just apply control rotation $R_Y({\pi y_r'}/{2^{t-1}})$ to $|0\rangle$.
If $y'_{r,0}=1$, then we know $y_r'\in\{2^{t-1},2^t-y_r\}$. In this case, we apply $X$ gate to $|0\rangle$ first,
then apply control rotation $R_Y({\pi y_r'}/{2^{t-1}})$, finally apply $-X$ to the result.
The quantum circuit is shown in Figure \ref{fig_qnideala}.

If we consider $|\psi_i\rangle R_Y({\pi y_r}/{2^{t-1}})|0\rangle$, then
based on the fact $R_Z(\pi (2^t-y_i)/2^{t-1})|0\rangle = -X R_Z(\pi y_i/2^{t-1}) X$ and the above analysis,
we can generate $|\psi_i\rangle R_Z({\pi y_i}/{2^{t-1}}) R_Y({\pi y_r}/{2^{t-1}})|0\rangle$
by the quantum circuit of Figure \ref{fig_qnidealb}.

\begin{figure}[htbp]
\hspace{2em}\Qcircuit @C=1em @R=1em {
 \lstick{}& \qw{/\hspace{0.2em}^{t}}&\ctrl{2}  &\qw&\qw&\qw&\qw\\
  \lstick{}& \qw{\hspace{1.3em}/\hspace{0.2em}^{n+1}}&\qw  &\qw&\qw&\qw&\qw
  \inputgroup{1}{2}{.5em}{|\psi_r\rangle}\\
 \lstick{|0\rangle}& \qw & \gate{R_{Y_{f_r}}(t)} &\qw&\gate{R_{Z_{f_i}}(t)}&\gate{R_Z(-\frac{\pi}{2})}&\qw&|d\rangle\\
 &\qw{/\hspace{0.2em}^{t}}&\qw&\qw&\ctrl{-1}&\qw&\qw\\
 &\qw{\hspace{1.3em}/\hspace{0.2em}^{n+1}}&\qw&\qw&\qw&\qw&\qw
 \inputgroup{4}{5}{.5em}{|\psi_i\rangle}\\
}
\caption{The quantum neuron in the ideal case, where $|d\rangle = f(\langle x|w\rangle)$.}
\label{fig_qnideal}
\end{figure}
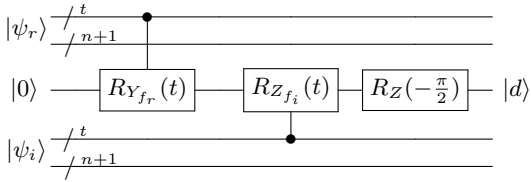

Finally, we conclude the above two procedures in Figure \ref{fig_qnideal} by adding $R_Z(-\pi/2)$ to generate $f(\langle x|w\rangle)$, where $R_{Y_{f_r}}(t)$
and $R_{Z_{f_i}}(t)$ are short for the control operators used in Figure \ref{fig_qnideala} and Figure \ref{fig_qnidealb} respectively.

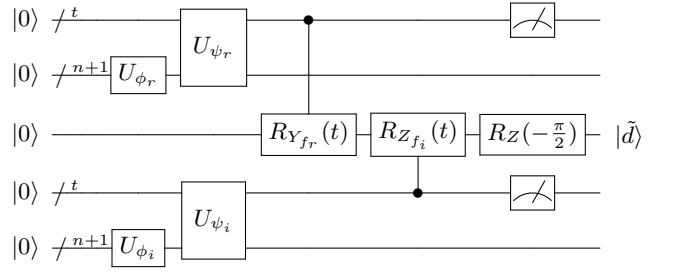
\begin{figure}[h]
\hspace{1.5em}\Qcircuit @C=0.6em @R=0.8em {
 \lstick{|0\rangle} &\qw{/\hspace{0.2em}^{t}}&\qw&\qw&\qw&\multigate{1}{U_{\psi_r}}&\ctrl{2}  &\qw&\meter&\qw&&\\
  \lstick{|0\rangle}& \qw{\hspace{1.3em}/\hspace{0.2em}^{n+1}}&\qw&\qw&\gate{U_{\phi_r}}&\ghost{U_{\psi_r}}&\qw  &\qw&\qw&\qw\\
 \lstick{|0\rangle}&\qw&\qw&\qw&\qw&\qw & \gate{R_{Y_{f_r}}(t)} &\gate{R_{Z_{f_i}}(t)}&\gate{R_Z(-\frac{\pi}{2})}&\qw&&|\tilde d\rangle\\
\lstick{|0\rangle}&\qw{/\hspace{0.2em}^{t}}&\qw&\qw&\qw&\multigate{1}{U_{\psi_i}}&\qw&\ctrl{-1}&\meter&\qw&&\\
\lstick{|0\rangle}&\qw{\hspace{1.3em}/\hspace{0.2em}^{n+1}}&\qw&\qw&\gate{U_{\phi_i}}&\ghost{U_{\psi_i}}&\qw&\qw&\qw&\qw\\
}
\caption{The quantum neuron in the general case.}
\label{fig_qngeneral}
\end{figure}

Generally, $\frac{\arccos-{\bm \Re}\langle x|w\rangle}{2\pi}$ and $\frac{\arccos-{\bm \Im}\langle x|w\rangle}{2\pi}$
cannot be written in binary form precisely. And $y_r,y_i$ only give approximates of them.
By introducing measurements to the original circuit, the quantum circuit
given in Figure \ref{fig_qngeneral} returns an approximate of $f(\langle x|w\rangle)$ with high probability.
For a detailed proof see Appendix \ref{the proof of th1}.

\begin{theorem}\label{thm_general}
Let $\ket{x}\in \C^{2^n}$ be a quantum state and
$\ket{w} \in (\C^2)^{\otimes n}$ be a product state.
Let $t=m+\lceil \log_{2}\left(2+\frac{1}{\sigma}\right)\rceil$ be the number of ancilla qubits used in quantum phase estimation,
where $\sigma\in(0,1)$ and $m\in\Z^{+}$.
Assume that $|\tilde{d}\rangle$ is the state obtained by the quantum circuit
given in Figure \ref{fig_qngeneral}. Then with success probability at least $1-\sigma$,
we have $\||\tilde{d}\rangle-|d\rangle\| \leq \pi/2^{m-1}$, where $|d\rangle=f(\langle x|w\rangle)$.
\end{theorem}

In Figure \ref{fig_qngeneral}, the purpose of performing measurements is simply to
convert the mixed state (\ref{eq_psi-exact}) in the ancillary registers into a pure state $|\tilde d\rangle$
that is close to $f(\langle x|w\rangle)$. However, it is unnecessary to record or store the measured results,
which makes it possible to perform quantum neurons without the classical resources.

One thing worth noting is that the quantum neuron model defined by Fig. \ref{fig_qngeneral} can be used to analyze
quantum data with real number labels through analyzing the measured results $|\tilde d\rangle$.
More precisely, assume that $|\tilde d\rangle = p_0|0\rangle + p_1|1\rangle$ is the output of Fig. \ref{fig_qngeneral}.
By equation (\ref{activtion function}), if we perform measurements to $|\tilde d\rangle$, then we can estimate
\[
|p_1|^2 \approx \sin^2(\frac{\arccos-{\bm \Re}\langle x|w\rangle}{2}) = \frac{1 + {\bm \Re}\langle x|w\rangle}{2}.
\]
The probability $|p_1|^2$ characterizes the closeness between $|\tilde d\rangle$
and $|1\rangle$. It can be viewed as the label of the input state $|x\rangle$.
Note that to solve the classification problems  by classical neural networks, we need to
calculate a function of the inner product between the input and the weight.
However, this inner product is already include in $|p_1|^2$.
Thus classical classification problems can also be solved by quantum neuron.
Especially for binary classification problems, we can simply define the label of $|x\rangle$
as a quantum state, e.g. $|0\rangle$ or $|1\rangle$.

\section{Construction of the quantum neural network}\label{sec_qnn}

The classical feed-forward neural network has been used to process data to simulate unknown nonlinear functions
\cite{sanger1989optimal,bebis1994feed,huang2004extreme}. In this section we introduce a quantum feed-forward neural network
to accomplish a similar task.

Let $\mathcal M\triangleq\{|x^i\rangle:i=1,\ldots,q\}\subset \C^{2^n}$ be a quantum data set.
We want to apply some kind of quantum feed-forward neural network
to capture the property and structure of $\mathcal M$ theoretically.
More precisely, suppose that the information of $\mathcal M$ is included in an unknown function $F_0$ mapping $\mathcal M$
to a product state space with dimensions $2^s$, that is
\be\ba{rll} \vspace{.2cm}
F_0:\hspace{.5cm}   M~   &\longrightarrow& (\C^2)^{\otimes s} \\
          |x^i\rangle    &\longmapsto&     |d^i\rangle=|d_1^i,\ldots,d_s^i\rangle.
\ea\ee
Our purpose is to construct a neural network based on the
quantum neuron to simulate $F_0$ efficiently.

Let $|x\rangle\in\mathcal M$ be the input state and it is allowed to be entangled.
For convenience, we assume $|x\rangle$ a product state, that is $|x\rangle=|x_1,x_2,\ldots, x_n\rangle$.
The state $|x\rangle$ constitutes the input layer, i.e., the 0-th layer, of the quantum neural network.
We denote it as $|z^{(0)}\rangle$.
Suppose we have $K-1$ hidden layers and one output layer. The output layer is also known as  the $K$-th layer.
Denote the number of neurons in the $k$-th layer as $p_k$, where $k=1,\ldots,K$ and $p_K=s$.

\begin{figure}[h]
  \centering
  \includegraphics[width=8.5cm]{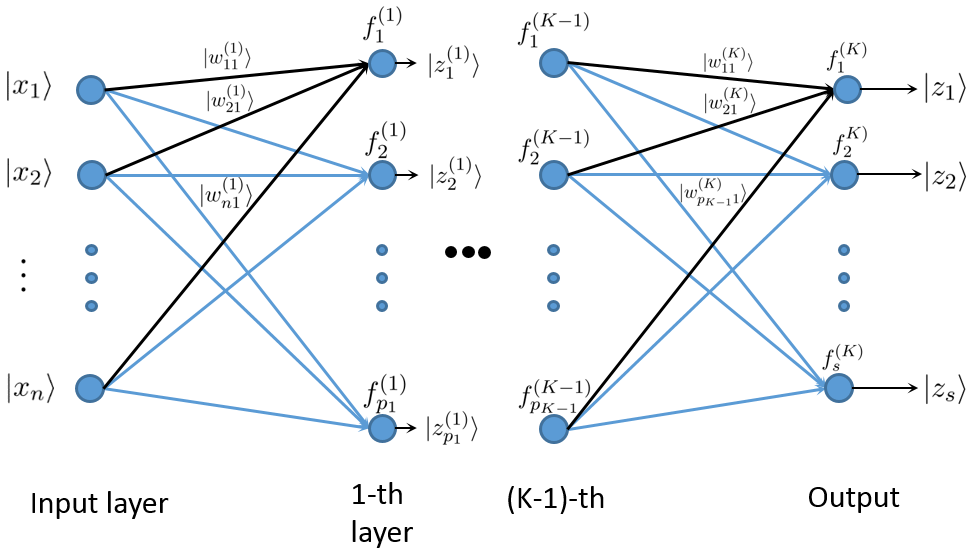}\\
  \caption{The quantum feed-forward neural network}
  \label{fig_qnn}
\end{figure}

For $k=1,\ldots,K$, the $j$-th neuron in the $k$-th and $(k-1)$-th layers are connected by an edge
with weight $|w_{ij}^{(k)}\rangle$, where $i=1,\dots,p_{k-1},j=1,\dots,p_{k}$.
The state of each neuron in the $k$-th layer is determined by the weights and the states of the $(k-1)$-th layer.
Thus, if we denote by
$|z_{j}^{(k)}\rangle$ as the state of the $j$-th neuron in the $k$-th layer, then
\be
|z_{j}^{(k)}\rangle=f_j^{(k)}(\langle z^{(k-1)}|w_j^{(k)}\rangle),
\ee
where $|z^{(k-1)}\rangle = |z_{1}^{(k-1)},\ldots, z_{p_{k-1}}^{(k-1)}\rangle$,
$|w_j^{(k)}\rangle=|w_{1j}^{(k)},\ldots, $ $w_{p_{k-1}j}^{(k)}\rangle$,
and $f_j^{(k)}$ is defined by equation (\ref{activtion function}).
Figure \ref{fig_qnn} shows the basic structure of quantum neural network.

\begin{exmp}
We set $n=2$, $p_1=2$, $K=2$ and $p_2=s=1$. In this case the quantum neural network and the corresponding
quantum circuit are shown in Figure \ref{fig_qnn_circuit} (a) and Figure \ref{fig_qnn_circuit} (b),
respectively.


\begin{figure*}[htbp]
\centering\begin{sideways}{
\subfigure[ ]{
\begin{minipage}{12cm}\centering
\includegraphics[width=12cm]{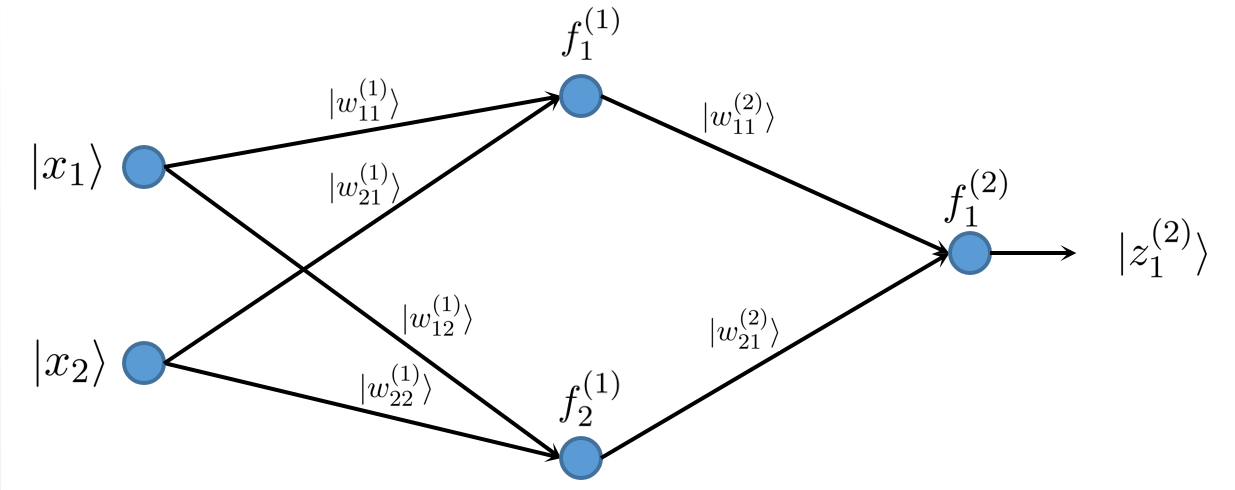}
\end{minipage}}}\end{sideways}
\begin{sideways}{
\subfigure[ ]{
\begin{minipage}{21.7cm}
\scalebox{0.63}[0.8]{
\Qcircuit @C=0.3em @R=0.5em {
\lstick{|0\rangle}&\qw{/\hspace{0.2em}^{t}}&\qw&\qw  &\qw &\qw &\qw&\qw&\qw&\qw&\gate{H^{\otimes t}}&\qw{\begin{matrix}|j\rangle\\
\hspace{1em}\end{matrix}} &\ctrl{1} &\gate{\rm {FT}^{\dag}} &\ctrl{16} &\qw&\qw&\qw&\qw&\qw&\qw&\qw&\qw&\qw&\qw&\qw&\qw&\qw&\meter \\
 \lstick{|0\rangle}& \qw &\gate{H}&\ctrl{1} &\qw &\ctrlo{1}&\ctrlo{1}&\ctrlo{2}&\ctrlo{2}&\ctrlo{1}&\qw&\gate{H}&\multigate{10}{\begin{matrix}&G_{r1}^{(2)j}\\& \\& \\&\\&\\&\\&\\&\\&\\&\\& \end{matrix}}&\qw&\qw&\qw&\qw&\qw&\qw&\qw&\qw&\qw&\qw&\qw&\qw&\qw&\qw\\
 \lstick{|0\rangle}& \qw &\qw&\multigate{1}{U_{w_{1}^{(2)}}\otimes U_{w_{2}^{(2)}}} &\qw &\gate{R_{Y_{f_{r1}^{(1)}}}(t)}&\gate{R_{Z_{f_{i1}^{(1)}}}(t)}\qw&\qw&\qw&\multigate{1}{R_Z^{\otimes 2}(-\frac{\pi}{2})}&\qw&\qw&\ghost{\begin{matrix}&G_{1r}^{(2)j}\\& \\& \\&\\&\\&\\&\\&\\&\\&\\& \end{matrix}}&\qw&\qw&\qw&\qw&\qw&\qw&\qw&\qw&\qw&\qw&\qw&\qw&\qw&\qw\\
   \lstick{|0\rangle}& \qw&\qw&\ghost{U_{w_{1}^{(1)}}\otimes U_{w_{2}^{(1)}}} &\qw &\qw&\qw &\gate{R_{Y_{f_{r2}^{(1)}}}(t)}&\gate{R_{Z_{f_{i2}^{(1)}}}(t)}\qw&\ghost{R_Z^{\otimes 2}(-\frac{\pi}{2})}&\qw&\qw&\ghost{\begin{matrix}&G_{1r}^{(2)j}\\& \\& \\&\\&\\&\\&\\&\\&\\&\\& \end{matrix}}&\qw&\qw&\qw&\qw&\qw&\qw&\qw&\qw&\qw&\qw&\qw&\qw&\qw&\qw\\
\lstick{|0\rangle}& \qw{/\hspace{0.2em}^{t}}&\qw&\qw  &\multigate{1}{U_{\Psi_{r1}^{(1)}}}&\ctrl{-2}&\qw&\qw&\qw&\qw&\qw&\qw&\ghost{\begin{matrix}&G_{i1}^{(2)j}\\& \\& \\&\\&\\&\\&\\&\\&\\&\\& \end{matrix}}&\qw&\qw&\ctrl{8}&\qw&\qw&\qw&\qw&\qw&\qw&\multigate{10}{\begin{matrix}& \\&\\& \\& \\&\\&\\&\\&\\&\\&\\&\\&G_{i1}^{(2)j} \end{matrix}}&\qw&\qw&\qw&\qw&\meter\\
  \lstick{|0\rangle}& \qw{/\hspace{0.2em}^{3}}&\qw&\gate{U_{\Phi_{r1}^{(1)}}}  &\ghost{U_{\Psi_{r1}^{(1)}}}&\qw&\qw&\qw&\qw&\qw&\qw&\qw&\ghost{\begin{matrix}&G_{1r}^{(2)j}\\& \\& \\&\\&\\&\\&\\&\\&\\&\\& \end{matrix}}&\qw&\qw&\qw&\qw&\qw&\qw&\qw&\qw&\qw&\ghost{\begin{matrix}&G_{1r}^{(2)j}\\& \\& \\&\\&\\&\\&\\&\\&\\&\\& \end{matrix}}&\qw&\qw&\qw&\qw\\
 \lstick{|0\rangle}& \qw{/\hspace{0.2em}^{t}}&\qw&\qw  &\multigate{1}{U_{\Psi_{i1}^{(1)}}}&\qw&\ctrl{-4}&\qw&\qw&\qw&\qw&\qw&\ghost{\begin{matrix}&G_{1r}^{(2)j}\\& \\& \\&\\&\\&\\&\\&\\&\\&\\& \end{matrix}}&\qw&\qw&\qw&\ctrl{6}&\qw&\qw&\qw&\qw&\qw&\ghost{\begin{matrix}&G_{1r}^{(2)j}\\& \\& \\&\\&\\&\\&\\&\\&\\&\\& \end{matrix}}&\qw&\qw&\qw&\qw&\meter\\
  \lstick{|0\rangle}& \qw{/\hspace{0.2em}^{3}}&\qw&\gate{U_{\Phi_{i1}^{(1)}}}  &\ghost{U_{\Psi_{i1}^{(1)}}}&\qw&\qw&\qw&\qw&\qw&\qw&\qw&\ghost{\begin{matrix}&G_{1r}^{(2)j}\\& \\& \\&\\&\\&\\&\\&\\&\\&\\& \end{matrix}}&\qw&\qw&\qw&\qw&\qw&\qw&\qw&\qw&\qw&\ghost{\begin{matrix}&G_{1r}^{(2)j}\\& \\& \\&\\&\\&\\&\\&\\&\\&\\& \end{matrix}}&\qw&\qw&\qw&\qw\\
 \lstick{|0\rangle}& \qw{/\hspace{0.2em}^{t}}&\qw&\qw  &\multigate{1}{U_{\Psi_{r2}^{(1)}}}&\qw&\qw&\ctrl{-5}&\qw&\qw&\qw&\qw&\ghost{\begin{matrix}&G_{1r}^{(2)j}\\& \\& \\&\\&\\&\\&\\&\\&\\&\\& \end{matrix}}&\qw&\qw&\qw&\qw&\ctrl{5}&\qw&\qw&\qw&\qw&\ghost{\begin{matrix}&G_{1r}^{(2)j}\\& \\& \\&\\&\\&\\&\\&\\&\\&\\& \end{matrix}}&\qw&\qw&\qw&\qw&\meter\\
  \lstick{|0\rangle}& \qw{/\hspace{0.2em}^{3}}&\qw&\gate{U_{\Phi_{r2}^{(1)}}}  &\ghost{U_{\Psi_{r2}^{(1)}}}&\qw&\qw&\qw&\qw&\qw&\qw&\qw&\ghost{\begin{matrix}&G_{1r}^{(2)j}\\& \\& \\&\\&\\&\\&\\&\\&\\&\\& \end{matrix}}&\qw&\qw&\qw&\qw&\qw&\qw&\qw&\qw&\qw&\ghost{\begin{matrix}&G_{1r}^{(2)j}\\& \\& \\&\\&\\&\\&\\&\\&\\&\\& \end{matrix}}&\qw&\qw&\qw&\qw\\
 \lstick{|0\rangle}& \qw{/\hspace{0.2em}^{t}}&\qw&\qw  &\multigate{1}{U_{\Psi_{i2}^{(1)}}}&\qw&\qw&\qw&\ctrl{-7}&\qw&\qw&\qw&\ghost{\begin{matrix}&G_{1r}^{(2)j}\\& \\& \\&\\&\\&\\&\\&\\&\\&\\& \end{matrix}}&\qw&\qw&\qw&\qw&\qw&\ctrl{3}&\qw&\qw&\qw&\ghost{\begin{matrix}&G_{1r}^{(2)j}\\& \\& \\&\\&\\&\\&\\&\\&\\&\\& \end{matrix}}&\qw&\qw&\qw&\qw&\meter\\
  \lstick{|0\rangle}& \qw{/\hspace{0.2em}^{3}}&\qw&\gate{U_{\Phi_{i2}^{(1)}}}  &\ghost{U_{\Psi_{i2}^{(1)}}}&\qw&\qw&\qw&\qw&\qw&\qw&\qw&\ghost{\begin{matrix}&G_{1r}^{(2)j}\\& \\& \\&\\&\\&\\&\\&\\&\\&\\& \end{matrix}}&\qw&\qw&\qw&\qw&\qw&\qw&\qw&\qw&\qw&\ghost{\begin{matrix}&G_{1r}^{(2)j}\\& \\& \\&\\&\\&\\&\\&\\&\\&\\& \end{matrix}}&\qw&\qw&\qw&\qw\\
\lstick{|0\rangle}& \qw&\qw&\multigate{1}{U_{w_{1}^{(2)}}\otimes U_{w_{2}^{(2)}}} &\qw&\qw&\qw&\qw&\qw&\qw&\qw&\qw&\qw&\qw&\qw &\gate{R_{Y_{f_{r1}^{(1)}}}(t)}&\gate{R_{Z_{f_{i1}^{(1)}}}(t)}&\qw&\qw &\multigate{1}{R_Z^{\otimes 2}(-\frac{\pi}{2})}&\qw&\qw&\ghost{\begin{matrix}&G_{1r}^{(2)j}\\& \\& \\&\\&\\&\\&\\&\\&\\&\\& \end{matrix}}&\qw&\qw&\qw&\qw\\
\lstick{|0\rangle}& \qw &\qw&\ghost{U_{w_{1}^{(1)}}\otimes U_{w_{2}^{(1)}}} &\qw&\qw&\qw&\qw&\qw&\qw&\qw&\qw&\qw&\qw&\qw &\qw&\qw&\gate{R_{Y_{f_{r2}^{(1)}}}(t)}&\gate{R_{Z_{f_{i2}^{(1)}}}(t)}&\ghost{R_Z^{\otimes 2}(-\frac{\pi}{2})}&\qw&\qw&\ghost{\begin{matrix}&G_{1r}^{(2)j}\\& \\& \\&\\&\\&\\&\\&\\&\\&\\& \end{matrix}}&\qw&\qw&\qw&\qw\\
\lstick{|0\rangle}& \qw &\gate{H}&\ctrl{-1} &\qw&\qw&\qw&\qw&\qw&\qw&\qw&\qw&\qw&\qw&\qw &\ctrlo{-2}&\ctrlo{-2}&\ctrlo{-1}&\ctrlo{-1}&\ctrlo{-1}&\gate{R_Z(-\frac{\pi}{2})}&\gate{H}&\ghost{\begin{matrix}&G_{1r}^{(2)j}\\& \\& \\&\\&\\&\\&\\&\\&\\&\\& \end{matrix}}&\qw&\qw&\qw&\qw\\
\lstick{|0\rangle}&\qw{/\hspace{0.2em}^{t}}&\qw&\qw  &\qw&\qw&\qw &\qw&\qw&\qw&\qw&\qw&\qw&\qw&\qw&\qw&\qw&\qw&\qw &\gate{H^{\otimes t}}&\qw{\begin{matrix}&\hspace{-6.5em}|j\rangle \\&
 \end{matrix}} &\qw&\ctrl{-1} &\gate{\rm {FT}^{\dag}} &\ctrl{1} &\qw&\qw&\qw&\meter\\
\lstick{|0\rangle}& \qw&\qw&\qw  &\qw &\qw &\qw&\qw&\qw&\qw &\qw&\qw &\qw &\qw &\gate{R_{Y_{f_{r1}^{(2)}}}(t)} &\qw&\qw&\qw&\qw&\qw&\qw&\qw&\qw&\qw&\gate{R_{Z_{f_{i1}^{(2)}}}(t)}&\gate{R_Z(-\frac{\pi}{2})}&\qw&\qw&\qw&\hspace{1em}|z_{1}^{(2)}\rangle
}}\\
\end{minipage}
}}\end{sideways}\caption{\textbf{Construction of the quantum feedforward neural network.} Here the input states are $|x_1\rangle$ and $|x_2\rangle$, and the output state is $|z_1^{(2)}\rangle$. (a) \textbf{Quantum neural network model with 3 neurons.} (b) \textbf{The quantum neural network represented by a circuit.} The transformations $G_{r1^{(2)}}$ and $G_{i1}^{(2)}$ are controlled by 4t qubits compared to before; see Figure \ref{fig_Gr}.   \label{fig_qnn_circuit}}
\end{figure*}
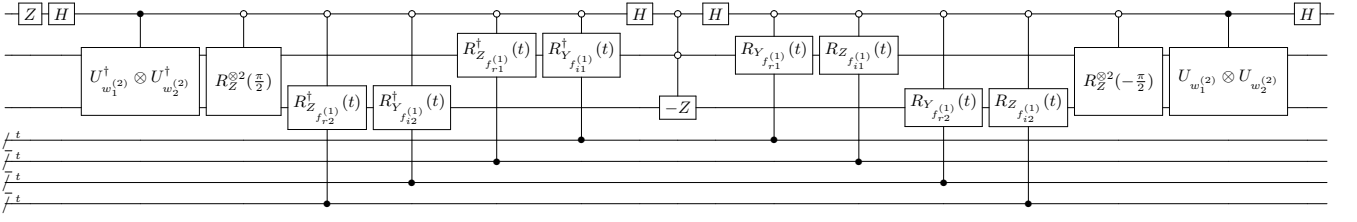
\begin{figure*}[tbp]
\scalebox{0.7}[0.7]{
\hspace{0.02em}\Qcircuit @C=0.4em @R=.5em {
&\qw&\gate{Z}&\gate{H} &\ctrl{1}&\ctrlo{1}&\ctrlo{2}&\ctrlo{2}&\ctrlo{1}&\ctrlo{1}&\gate{H}&\ctrlo{1}&\gate{H}&\ctrlo{1}&\ctrlo{1}&\ctrlo{2}&\ctrlo{2}&\ctrlo{1}&\ctrl{1}&\gate{H}&\qw&\qw\\
&\qw&\qw&\qw&\multigate{1}{U_{w_{1}^{(2)}}^{\dag}\otimes U_{w_{2}^{(2)}}^{\dag}} &\multigate{1}{R_Z^{\otimes 2}(\frac{\pi}{2})}&\qw&\qw&\gate{R_{Z_{f_{r1}^{(1)}}}^{\dag}(t)}&\gate{R_{Y_{f_{i1}^{(1)}}}^{\dag}(t)}&\qw&\ctrlo{1}&\qw&\gate{R_{Y_{f_{r1}^{(1)}}}(t)}&\gate{R_{Z_{f_{i1}^{(1)}}}(t)}&\qw&\qw&\multigate{1}{R_Z^{\otimes 2}(-\frac{\pi}{2})}&\multigate{1}{U_{w_{1}^{(2)}}\otimes U_{w_{2}^{(2)}}}&\qw&\qw\\
& \qw&\qw&\qw&\ghost{U_{w_{1}^{(2)}}^{\dag}\otimes U_{w_{2}^{(2)}}^{\dag}} &\ghost{R_Z^{\otimes 2}(\frac{\pi}{2})}&\gate{R_{Z_{f_{r2}^{(1)}}}^{\dag}(t)}&\gate{R_{Y_{f_{i2}^{(1)}}}^{\dag}(t)}&\qw&\qw&\qw&\gate{-Z}&\qw&\qw&\qw&\gate{R_{Y_{f_{r2}^{(1)}}}(t)}&\gate{R_{Z_{f_{i2}^{(1)}}}(t)}&\ghost{R_Z^{\otimes 2}(-\frac{\pi}{2})}&\ghost{U_{w_{1}^{(2)}}\otimes U_{w_{2}^{(2)}}}&\qw&\qw\\
&\qw{/\hspace{0.3em}^t}&\qw&\qw&\qw&\qw&\qw&\qw&\qw&\ctrl{-2}&\qw&\qw&\qw&\ctrl{-2}&\qw&\qw&\qw&\qw&\qw&\qw&\qw\\
&\qw\\
&\qw{/\hspace{0.3em}^t}&\qw&\qw&\qw&\qw&\qw&\qw&\ctrl{-4}&\qw&\qw&\qw&\qw&\qw&\ctrl{-4}&\qw&\qw&\qw&\qw&\qw&\qw\\
&\qw\\
&\qw{/\hspace{0.3em}^t}&\qw&\qw&\qw&\qw&\qw&\ctrl{-5}&\qw&\qw&\qw&\qw&\qw&\qw&\qw&\ctrl{-5}&\qw&\qw&\qw&\qw&\qw\\
&\qw\\
&\qw{/\hspace{0.3em}^t}&\qw&\qw&\qw&\qw&\ctrl{-7}&\qw&\qw&\qw&\qw&\qw&\qw&\qw&\qw&\qw&\ctrl{-7}&\qw&\qw&\qw&\qw
}}\caption{\textbf{The transformation $G_{r1}^{(2)}$. }\label{fig_Gr}}
\end{figure*}
\end{exmp}

In the construction of circuits we use the strategy of postponing  measurement. To be specific, we postpone the measured process of each neuron in all hidden layers until the last layer. In Figure \ref{fig_qnn_circuit} (b) we postpone $4t$ measured results in the first layer.

The strategy of postponing measurement is necessary. Suppose we want to get
the output state of the neuron in hidden layers, we need to measure the corresponding
qubits to convert the mixed state to a random pure state. Without postponing measurement
we cannot use the method of swap test to get the subsequent output states, which means
the neural network is interrupted. This implies that the intermediate state is unreadable
in the quantum neural network and we do not care about the state of hidden layer neurons naturally.

In this quantum neural network, we give quantitative estimations of
success probability and fidelity for the output state.
Its proof is presented in Appendix \ref{the proof of thm2}.

\begin{theorem}\label{thm_K layers}
Given a quantum neural network defined in Figure \ref{fig_qnn}.
Suppose the number of the neurons in the $k$-th layer is $p_k$.
Let $p=\max\{p_1, \dots,p_K\}$,
$\epsilon\in(0,1)$ and $\sigma\in(0,1)$. Set
$m=\lceil \log_2[(\frac{2\pi^2p^2}{\epsilon})^{2^{K-1}}\pi] \rceil+1$ and
$t=m+\lceil \log_2(2+\frac{Kp}{\sigma}) \rceil$.
Then with success probability at least $1-\sigma$ we have the fidelity
\be
\left\| |z^{(K)}\rangle - |\tilde{z}^{(K)}\rangle \right\| \leq \epsilon.
\ee
\end{theorem}


\section{Training process}\label{sec_training process}

In this section, we introduce the training process of the proposed quantum neural network. We transform the quantum neural network into a quantum circuit containing parameters to be optimized.  The training process of parameterized quantum circuits has been used in many quantum algorithms \cite{peruzzo2014variational,farhi2014quantum,mitarai2018quantum}.

Suppose the quantum neural network has $n$ neurons in the input layer and has $s$ neurons in the output layer.
In training process, we choose the mean square loss
\begin{align}\label{eq_loss1}
\notag\mathcal L(\mathcal M,\mathcal W)
=& \frac{1}{q} \sum_{i=1}^q \left| |z^i\rangle-|d^i\rangle \right|^2 \\
=& \frac{1}{q} \sum_{i=1}^q (2-2{\bf Re}\langle z^i|d^i\rangle).
\end{align}
Each input state $|x^i\rangle\in\mathcal M$ has a fixed label $|d^i\rangle=|d^i_1,\dots,d^i_s\rangle$.
Each output state $|z^i\rangle=|z^i_1,\dots, z^i_s\rangle$ produced by quantum circuits can be closed to $|\tilde z^i\rangle$ with high success probability according to theorem 2, where $|\tilde z^i\rangle$ is the ideal output state decided by all the weights $|w_{j}^{(k)}\rangle$ and the activation function $f$ defined in expression \eqref{activtion function}.

Our goal is to find a set $\mathcal W\triangleq\{|w_{j}^{(k)}\rangle:k=1,\dots,K,j=1,\ldots,p_k\}$ of
weight states such that they minimize the mean square loss.

Since $|w_{j}^{(k)}\rangle=|w_{1j}^{(k)},\ldots,w_{p_{k-1}j}^{(k)}\rangle$ is a product state, we assume that
\be
|w_{ij}^{(k)}\rangle = e^{\bm i \delta_{ijk}} R_Z(\gamma_{ijk}) R_Y(\beta_{ijk}) |0\rangle
\ee
for some parameters $\beta_{ijk},\gamma_{ijk},\delta_{ijk}\in[0,2\pi)$ to be tuned. Denote the parameter vector by $\theta=(\theta_1,\dots,\theta_L)^{\rm T}$, where $\theta_i\in\{\beta_{ijk},\gamma_{ijk},\delta_{ijk}\}$ and $L=|\{\beta_{ijk},\gamma_{ijk},\delta_{ijk}\}|$.
As the Figure \ref{fig_qngeneral} and the Figure \ref{fig_qnn_circuit}(b), the output state $|z^i\rangle$ always can be obtained by performing a unitary transformation, denoted by $U^i(\theta)$, to the initial state $|0\rangle$ and adding some measurements.
Let $|Z^i\rangle= U^i(\theta)|0\rangle$, then the output state $|z^i\rangle$ is decided by the parameter vector $\theta$ and measurement results.
Denote the map from $|0\rangle$ to $|z^i\rangle$ by $F^i(\theta)$.
Thus, $\mathcal{L}$ can be viewed as a function of $\theta$.

We explain the training process as follows.

{\bf Step 1. Initial value selection.} Randomly try the initial parameter vector $\theta$ and choose the optimal parameter denoted by $\theta^{(0)}$ such that the value of $\mathcal L$ is minimum.

 The value of ${\bf Re}\langle z^i|d^i\rangle$ for each vector $\theta$ can be obtained by reusing quantum swap test. Then classically calculate and compare the different values of $\mathcal L(\theta)$ to obtain the optimal initial value.

 {\bf Step 2. Iteration process.} We use the gradient descendent method. In the $(i+1)$-th step,
 \begin{align}\label{eq_itera}
 \theta^{(i+1)}_l=\theta^{(i)}_l-\eta\frac{\partial \mathcal L}{\partial \theta_l},
 \end{align}
 where $\eta$ is an adjustable positive step length and $l=1,\dots,L$.
Combining the expressions \eqref{eq_loss1}\eqref{eq_itera}, we can use the quantum-classical hybrid method to acquire the gradient.
 \begin{align}\label{eq_par L}
\frac{\partial \mathcal L}{\partial \theta_l}=&
-\frac{2}{q}\sum_{i=1}^q\frac{{\bf Re}\partial\langle d^i|z^i\rangle}{\partial \theta_l},\\
\frac{\partial|z^i\rangle}{\partial \theta_l}=&
\frac{\partial F^i}{\partial \theta_l}|0\rangle.
\end{align}

The partial derivative of $F^i$ can be obtained by firstly deriving the partial derivative of $U^k$ and then add the corresponding measurements.

To be specific, theoretically for arbitrary unitary transformation, it always can be represented by the basic unitary gates: the single particle rotation gates and the control X gates.
For example, if
$U^i=(R_X(2g_1(\theta_l))\otimes R_Z(2g_2(\theta_l)))(CNOT)(I\otimes R_Z(2g_3(\theta_{l'})))$, where $g_j(\theta_l)\in[0,2\pi)$ denotes the rotation angle for the single particle gate in the form by the basic unitary gates, $j=1,2,3$.
\begin{align}\label{eq_par_Ui}
&\frac{\partial U^i|0\rangle}{\partial \theta_l}
=-g'_1(\theta_l)|\bm i (X\otimes I)U^i|0\rangle-g'_2(\theta_l)|\bm i (I\otimes Z) U^i|0\rangle.
\end{align}
As expression \eqref{eq_par_Ui}, we can construct the quantum circuit for the unitary transformation $\bm i (X\otimes I)U^i$ and $\bm i (I\otimes Z) U^i$ , respectively.
Then measure and record the corresponding registers, collapsing $\bm i (X\otimes I)U^i|0\rangle$ and $\bm i (I\otimes Z) U^i|0\rangle$ to the states denoted by $|z_{p1}^i\rangle$ and $|z_{p2}^i\rangle$, respectively. At last, we use swap test to get the value of ${\bf Re}\langle d^i|z^i_{p_1}\rangle$ and ${\bf Re}\langle d^i|z^i_{p_2}\rangle$ and calculate the gradient of $\mathcal L$ by
\begin{align*}
\frac{{\bf Re}\partial\langle d^i|z^i\rangle}{\partial \theta_l}
=-g'_1(\theta_l){\bf Re}\langle d^i|z^i_{p_1}\rangle-g'_2(\theta_l){\bf Re}\langle d^i|z^i_{p_2}\rangle.
\end{align*}

\section{Numerical experiment: classification on a checkerboard}\label{sec_numer}

In this section, we numerically validate our model with the following checkerboard classification task.

Consider a product state $R_{Y}\left(\theta_{1}\right)|0\rangle \otimes R_{Y}\left(\theta_{2}\right)|0\rangle$. This state has two parameters $\theta_1, \theta_2 \in [0, 2\pi)$, which forms a square area $C\triangleq\{[0,2\pi)\times[0,2\pi)\}$. Now, suppose we divide the square $C$ into a $2\times 2$ checkerboard with two disjoint parts:

\begin{equation}
\begin{aligned}
C_{0} &=\{[0, \pi) \times[0, \pi)\} \cup\{[\pi, 2 \pi) \times[\pi, 2 \pi)\} \\
C_{1} &=\{[0, \pi) \times[\pi, 2 \pi)\} \cup\{[\pi, 2 \pi) \times[0, \pi)\}
\end{aligned}
\end{equation}

The task is to classify whether the input quantum state is in the region $C_0$ (labeled by $|0\rangle$) or $C_1$ (labeled by $|1\rangle$). To this end, we constructed a 4-layered quantum neural network: the input layer has two neurons corresponding to the input state, followed by two hidden layers with $8$ neurons in each of them, and the output layer has one neuron.

\begin{figure}[h]
  \centering
  \scalebox{0.21}{\includegraphics{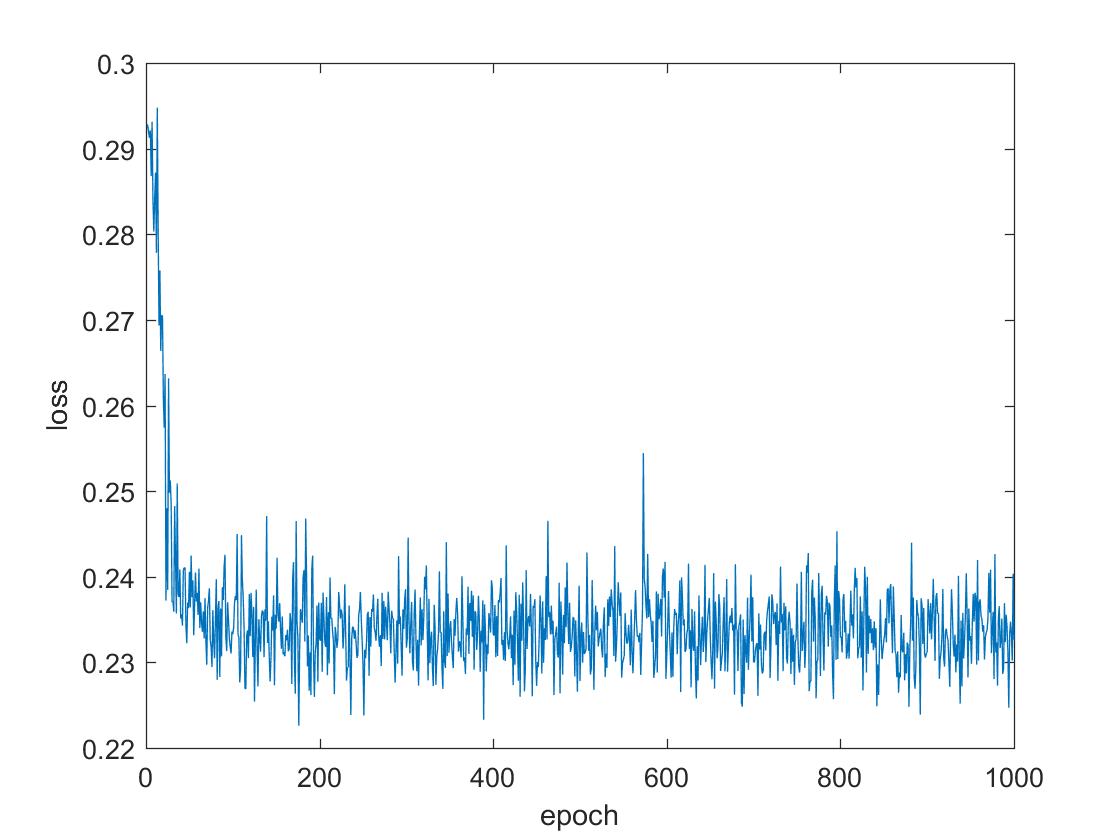}}
  \caption{The learning curve\label{fig_qnn_loss}}
\end{figure}

During the training process, we randomly generated $10^5$ data points, and applied the stochastic gradient descent algorithm to minimize the loss function defined in  equation \eqref{eq_loss1}. In this numerical experiment, since the vector forms of samples are known, we calculated the loss function and the gradient in the classical way. The learning curve is shown in Figure \ref{fig_qnn_loss}, from which we can see that the loss converged to about $0.23$.

\begin{figure}[h]
  \centering
  \scalebox{0.18}{\includegraphics{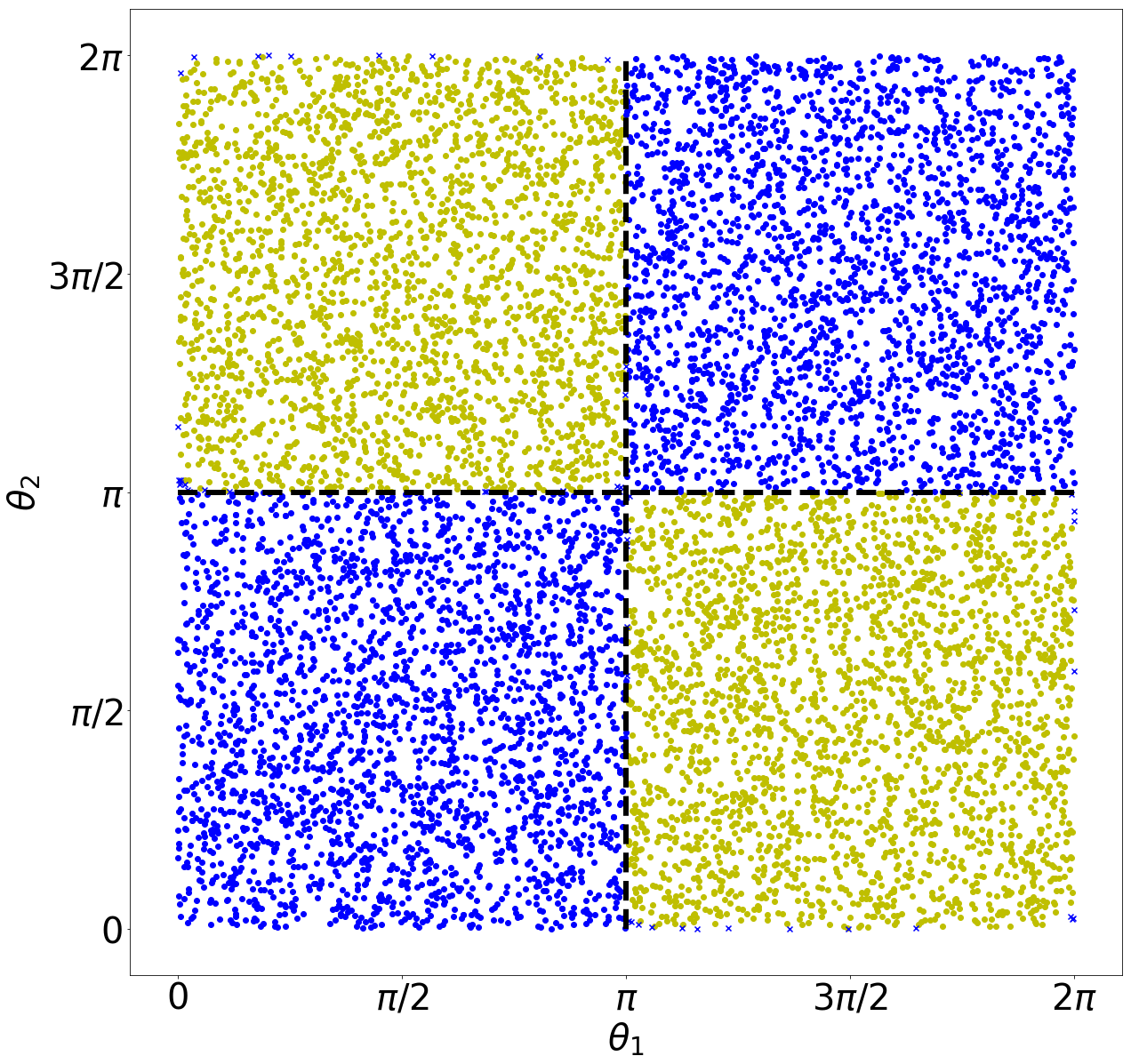}}
  \caption{The testing result:
  The correct predictions are represented with dots, and the incorrect predictions are labeled with
  crosses.}\label{fig_qnn_plots}
\end{figure}

After training, we further generated $10, 000$ samples to test our quantum neural network. The result is plotted in Figure \ref{fig_qnn_plots}, in which the classification accuracy achieved $99.25\%$.

\section{Conclusions}
\label{sec_conclusion}

The quantum neural network is introduced and its explicit expression is obtained.
The validity of the training process of neural network is proved theoretically.
The numerical example illustrates the potential of this model. Although there
exists the process of measurement, we do not need to record or store any measured
result, which means performing the quantum neural networks do not cost the resources for classical calculations.

This proposed quantum neural network includes some situations of classical neural
network, where the weights constitute a
vector belonging to the product state space. And it can be used to process both
quantum data with classical labels directly and classical data with classical labels by using state preparation.

A possible future research topic is to generalize the form of the weights in each
layer, such as $|w_{j}^{(k)}\rangle$ is not limited to the product state. One can
also generalize the activation operator $f$, which still retains the validity or
to generalize output state of neural network into a entangled state.

\section*{acknowledgements}

This work was supported by the National Key Research and Development
Program of China (Grant No. 2016YFA0301700) and the Anhui Initiative in
Quantum Information Technologies (Grant No. AHY080000).

\newpage
\section*{Appendix}

\renewcommand\thesection{A}
\section{The proof of Theorem 1}\label{the proof of th1}
\renewcommand{\theequation}{A\arabic{equation}}
\setcounter{equation}{0}
\proof
Denote $y_r^{\prime } = \lfloor{2^t\theta_r}/{\pi}\rfloor$ and $y_r^{\prime\prime } =2^t-y_r^{\prime } $
(see equation (\ref{eq_phi_theta_0u1v}) for the meaning of $\theta_r$).
By quantum phase estimation (see \cite{nielsen2002quantum}), $\forall \sigma'\in(0,1)$ we can choose
$t=m+\lceil \log_2(2+\frac{1}{2\sigma'})$ and
approximate ${\theta}_r/{\pi}$
to precision $2^{-m}$ with probability at leats $1-\sigma'$,
thus the exact form of the state $|\psi_r\rangle$ in equation \eqref{eq_psi} should be
\be\ba{lll} \vspace{.2cm} \label{eq_psi-exact}
& & \ds
\frac{-\bm i}{\sqrt{2}}\Bigg[ e^{\bm i \theta} \Big(
\sum_{ \substack{\tilde{y}_r':|\tilde{y}_r'-y_r^{\prime } | \\ \leq 2^{t-m}-1} } \beta_{\tilde{y}_r'} |\tilde{y}_r'\rangle
+\sum_{ \substack{\hat{y}_r':|\hat{y}_r'-y_r^{\prime } | \\ > 2^{t-m}-1} } \beta_{\hat{y}_r'} |\hat{y}_r'\rangle \Big)
|w_+\rangle \\
&-& \ds e^{-\bm i \theta} \Big(
\sum_{ \substack{\tilde{y}_r'':|\tilde{y}_r''-y_r^{\prime\prime } | \\ \leq 2^{t-m}-1} } \beta_{\tilde{y}_r''} |\tilde{y}_r''\rangle
+\sum_{ \substack{\hat{y}_r'':|\hat{y}_r''-y_r^{\prime\prime } | \\ > 2^{t-m}-1} }  \beta_{\hat{y}_r''} |\hat{y}_r''\rangle \Big)
|w_-\rangle \Bigg].
\ea\ee
Moreover,
\[
\ds \sum_{ \substack{\tilde{y}_r':|\tilde{y}_r'-y_r^{\prime } | \\ \leq 2^{t-m}-1} } \frac{|\beta_{\tilde{y}_r'}|^2}{2}  \geq \ds \frac{1-\sigma'}{2}, \hspace{.5cm}
\ds \sum_{ \substack{\tilde{y}_r'':|\tilde{y}_r''-y_r^{\prime\prime } | \\ \leq 2^{t-m}-1} } \frac{|\beta_{\tilde{y}_r''}|^2}{2} \geq \ds \frac{1-\sigma'}{2}.
\]

In $|\psi_r\rangle$, all $\tilde{y}_r'$ provide $2^{-m}$ approximates of $\theta_r/\pi$,
i.e., $|\tilde{y}_r'/2^t - \theta_r/\pi|\leq 2^{-m}$.
We also have $\tilde{y}_r'' = 2^t - \tilde{y}_r'$.
Apply control rotation shown in Figure \ref{fig_qnideala} to $|\psi_r\rangle|0\rangle$,
then with probability at least $1-\sigma'$, we  get
\be
|\tilde{d}_r'\rangle = R_Y(\tilde{y}_r'{\pi}/{2^{t-1}})|0\rangle
\ee
in the third register.
Denote the the angle between $|\tilde{d}_r'\rangle$ and $|d_r\rangle:=R_Y(2\theta_r)|0\rangle$ in Bloch sphere as $\eta_r$, then
\be \label{eq_eta_2}
\eta_r =  \Big| \tilde{y}_r' - \frac{2^t\theta_r}{\pi} \Big| \frac{\pi}{2^{t-1}} \leq \frac{\pi}{2^{m-1}}.
 \ee
Thus,
\be
\||\tilde{d}_r'\rangle - |d_r\rangle\| \leq \sqrt{2-2\cos(\eta_r/2)} = 2\sin({\eta_r}/{4}) \leq \pi/2^m.
\ee

Similarly, with probability at least $1-\sigma'$, we can obtain a
$\tilde{y}_i'$ such that $|\tilde{y}_i'/2^t - \theta_i/\pi|\leq 2^{-m}$.
By definition,
\[\ba{lll} \vspace{.2cm}
|\tilde{d}\rangle &=& R_Z(-\pi/2)R_Z(\tilde{y}_i'\pi/2^{t-1})R_Y(\tilde{y}_r'\pi/2^{t-1})|0\rangle, \\
|d\rangle &=& R_Z(-\pi/2)R_Z(2\theta_i)R_Y(2\theta_r)|0\rangle.
\ea\]
Therefore,
\[\ba{lll} \vspace{.2cm}
&& \||\tilde{d}\rangle-|d\rangle\| \\  \vspace{.2cm}
&\leq& \|R_Z(\tilde{y}_i'\pi/2^{t-1}) (R_Y(\tilde{y}_r'\pi/2^{t-1})|0\rangle\\ \vspace{.2cm}
&&  \hfill -\, R_Z(\tilde{y}_i'\pi/2^{t-1})R_Y(2\theta_r)|0\rangle) \| \\ \vspace{.2cm}
&& + \, \|R_Z(\tilde{y}_i'\pi/2^{t-1})R_Y(2\theta_r)|0\rangle)\\ \vspace{0.2cm}
&& \hfill - \, R_Z(2\theta_i)R_Y(2\theta_r)|0\rangle\| \\ \vspace{.2cm}
&\leq& \||\tilde{d}_r'\rangle - |d_r\rangle\| + \|R_Z(\tilde{y}_i'\pi/2^{t-1}) - R_Z(2\theta_i)\| \\
&\leq& \pi/2^m + \pi/2^m = \pi/2^{m-1}.
\ea\]
The success probability is $(1-\sigma')^2>1-2\sigma'$. We choose $\sigma=2\sigma'\in(0,1)$ and $t=m+\lceil \log_2(2+\frac{1}{\sigma}) \rceil$.
\eproof

\renewcommand\thesection{B}
\section{The details of Theorem 2}\label{the proof of thm2}
\renewcommand{\theequation}{B\arabic{equation}}
\setcounter{equation}{0}

\begin{lemma}\label{lemma}
Assume that $|x\rangle = |x_1,\ldots,x_n\rangle,|\tilde{x}\rangle = |\tilde{x}_1,\ldots,\tilde{x}_n\rangle$,
where $\||x_i\rangle-|\tilde{x}_i\rangle\| \leq \epsilon$ for all $i$.
Assume that $|w\rangle = |w_1,\ldots,w_n\rangle$. Then

(1). $\||x\rangle - |\tilde{x}\rangle\| \leq n\epsilon$.

(2). Let $g(y)=\arccos(-y)$, $y\in[-1,1]$. $\forall \delta\in(0,2)$, if $|y_1-y_2|<\delta$, then
\begin{align*}
|g(y_1)-g(y_2)|\leq \pi\sqrt{\delta}/\sqrt 2.
\end{align*}

(3). Suppose that $y_r\pi/2^{t-1}$, $y_i\pi/2^{t-1}$, $\tilde{y}_r\pi/2^{t-1}$, $\tilde{y}_i\pi/2^{t-1}$ are $\pi/2^m$ approximates of
$2\theta_r=\arccos-{\bm \Re}\langle x|w\rangle,$ $2\theta_i=\arccos-{\bm \Im}\langle x|w\rangle,$
$\arccos-{\bm \Re}\langle \tilde{x}|w\rangle,$ $\arccos-{\bm \Im}\langle \tilde{x}|w\rangle$
respectively, then
\[\ba{lll} \vspace{.2cm}
|\tilde{d}\rangle &=& R_Z(-\pi/2)R_Z(\tilde{y}_i\pi/2^{t-1})R_Y(\tilde{y}_r\pi/2^{t-1})|0\rangle, \\
|d\rangle &=& R_Z(-\pi/2)R_Z(2\theta_i)R_Y(2\theta_r)|0\rangle,
\ea\]
satisfies
$\||\tilde{d}\rangle-|d\rangle\| \leq \pi/2^{m-1} + \pi\sqrt{n\epsilon}/\sqrt 2$.
\end{lemma}

\proof (1). We prove the result by induction. The result is true for $n=1$. Denote
$|x'\rangle=|x_2,\ldots,x_n\rangle$ and $|\tilde{x}'\rangle = |\tilde{x}_2,\ldots,\tilde{x}_n\rangle$, then by induction
$\| |x'\rangle - |\tilde{x}'\rangle\| \leq (n-1)\epsilon$. Thus
\[\ba{lll} \vspace{.2cm}
\||x\rangle - |\tilde{x}\rangle\| &\leq& \||x_1,x'\rangle - |\tilde{x}_1,x'\rangle\| + \| |\tilde{x}_1,x'\rangle - |\tilde{x}_1,\tilde{x}'\rangle\| \\
&\leq& n\epsilon.
\ea\]

(2). Since $|y_1-y_2|\leq \delta$, we have $|g(y_1)-g(y_2)|\leq \arccos(1-\delta)$. Note that $\cos(\pi\sqrt\delta/\sqrt2)<1-\delta$,
then $|g(y_1)-g(y_2)|\leq \pi\sqrt\delta/\sqrt2$.

(3). By (1), we have $\||x\rangle - |\tilde{x}\rangle\| \leq n\epsilon$, thus $| \langle w|x\rangle -  \langle w|\tilde{x}\rangle | \leq n\epsilon$.
Denote $2\tilde{\theta}_r=\arccos-{\bm \Re}\langle \tilde{x}|w\rangle,$ $2\tilde{\theta}_i=\arccos-{\bm \Im}\langle \tilde{x}|w\rangle$,
then by (2), $|\tilde{\theta}_r-\theta_r|\leq \pi\sqrt{n\epsilon}/2\sqrt{2},
|\tilde{\theta}_i-\theta_i|\leq \pi\sqrt{n\epsilon}/2\sqrt{2}$.
Set
$|d'\rangle = R_Z(-\pi/2)R_Z(2\tilde{\theta}_i)R_Y(2\tilde{\theta}_r)|0\rangle$, then
\[\ba{lll}  \vspace{.2cm}
\||\tilde{d}\rangle - |d\rangle\| &\leq& \||\tilde{d}\rangle - |d'\rangle\| + \||d\rangle - |d'\rangle\| \\
&\leq& \pi/2^{m-1} + \pi\sqrt{n\epsilon}/\sqrt2.
\ea\]
This completes the proof.
\eproof

Then combining lemma \ref{lemma} and theorem \ref{thm_general}, we give the proof of theorem \ref{thm_K layers}.

\proof
Denote the error to generate $|z^{(k)}\rangle$ as $\epsilon_k$, then $\epsilon_0=0$.
Assume that $m=\lceil \log_2({\pi}/{\delta}) \rceil+1$ for some $\delta$
such that $\delta \leq \frac{\pi}{\sqrt2}\sqrt{\epsilon_k}$ for all $k\geq 1$.

By lemma \ref{lemma}, $\epsilon_1 \leq p_1 \frac{\pi}{2^{m-1}} \leq p\delta$.
When $k\geq 2$ and $\epsilon_{k-1}\leq 2$,
\[\ba{lll} \vspace{.2cm}
\epsilon_k &\leq& p_k( \frac{\pi}{2^{m-1}} + \frac{\pi}{\sqrt2}\sqrt{\epsilon_{k-1}}) \\ \vspace{.2cm}
&\leq& p (\delta + \frac{\pi}{\sqrt{2}} \sqrt{\epsilon_{k-1}} ) \\
&\leq& \sqrt{2}\pi p \sqrt{\epsilon_{k-1}}.
\ea\]
Thus,
\[\ba{lll} \vspace{.2cm}
\epsilon_k &\leq& (\sqrt2\pi)^{1+\frac{1}{2}+\cdots+\frac{1}{2^{k-2}}}  p^{1+\frac{1}{2}+\cdots+\frac{1}{2^{k-1}}} \delta^{\frac{1}{2^{k-1}}} \\
&\leq& 2\pi^2 p^2 \delta^{\frac{1}{2^{k-1}}}.
\ea\]
Setting $\epsilon_K = \epsilon$ shows that $\delta = (\epsilon/2\pi^2 p^2)^{2^{K-1}}$.
And we can check that $\epsilon_k<2\pi^2p^2{\delta}^{\frac{1}{2^{k-1}}}<\epsilon<2$.

By theorem \ref{thm_general}, if $t=m+\log_2(2+\frac{1}{\sigma'})$ the success probability is
$(1-\sigma')^{Kp}$. Let
$\sigma=Kp\sigma'\in(0,1)$,
then
\begin{align*}
(1-\sigma')^{Kp}=(1-\frac{\sigma}{Kp})^{Kp}\geq 1-\sigma.
\end{align*}
\eproof

\bibliographystyle{unsrt}
{\small
 \bibliography{123}
}

%
%
%

%
%
%
%
%
%
%
%

\end{document}